\documentclass[twocolumn,showpacs,preprintnumbers,amsmath,amssymb]{revtex4}

\usepackage{graphicx}
\usepackage{dcolumn}
\usepackage{bm}
\usepackage{amsmath}

\begin{document}

\title{Effect of a Magnetic Field on the Dipole Echo in Glasses\\ with Nuclear Quadrupole Moments.}

\author{A.\,V. Shumilin}
\affiliation{A.F.Ioffe Institute, 194021 Saint Petersburg, Russia.}

\author{D.\,A. Parshin}

\affiliation{Saint Petersburg State Polytechnical University, 195251
Saint Petersburg, Russia }

\date{\today}

\begin{abstract}
The effect of a magnetic field on the dipole echo amplitude in glasses
at temperatures of about $10$\,mK caused by nonspherical nuclei with
electric quadrupole moments has been studied
theoretically. It has been shown that in this case, the two-level
systems (TLS's) that determine the glass properties at low
temperatures are transformed into more complicated multilevel
systems. These systems  have new properties as compared to usual
TLS's and, in particular, exhibit oscillations of electric dipole
echo amplitude in magnetic field. A general formula that describes
the echo amplitude in an arbitrary split TLS has been derived with
perturbation theory. Detailed analytic and numerical analysis of the
formula has been performed. The theory agrees qualitatively and
quantitatively well with experimental data.
\end{abstract}



\maketitle

\section{Introduction}

Glasses at temperature below $1$\,K display a number of universal
properties which are fundamentally different from the properties of
similar crystals. These properties are almost independent of the
chemical composition and are therefore due to the structure of
amorphous solid; more exactly due to absence of long-range order in
the glass \cite{AGVF}. One of such universal phenomena is a
two-pulse dipole echo.

The essence of the effect is the following. When a glass is
subjected to two short electromagnetic pulses at a frequency
$1$\,GHz separated by a time interval $\tau$, a response at the same
frequency may be detected at time $\tau$ after the second pulse.

According to the standard theory by Anderson, Halperin, Varma, and
Phillips \cite{AGVF}, all these properties are associated with the
existence of so-called two-level systems (TLS's) in the glass, which
are the groups of atoms tuneling between two minima of a double-well
potential. However, the microscopic nature of the TLS's in the
majority of glasses remains unknown.

An interesting phenomenon was discovered in 2002 \cite{albaze}. The
amplitude of the dipole echo in some nonmagnetic glasses exhibited a
strong nonmonotonic (oscillating) dependence on the magnetic field
even in weak fields (about $10$\,mT). Wurger, Fleischmann, and Enss
\cite{quadr} suggested that the nonmonotonic magnetic field
dependence of the dipole echo amplitude is caused by the presence of
nuclei with the electric quadrupole moment (or, equivalently, the
nuclear spin $J \ge 1$) in the glass.

This suggestion qualitatively agrees with the experimental data, in
particular, with the results of the recent measurements of the
dipole echo in glycerol ($\rm C_3H_8O_3$) \cite{glyc}. In these
experiments, it was shown that the replacement of hydrogen, which
has a nuclear spin $J = 1/2$ and, consequently, a zero quadrupole
moment, by deuterium ($J = 1$ and a nonzero quadrupole moment)
results in a more than an order of magnitude enhancement of the
magnetic-field dependence of the echo amplitude.

The magnetic-field dependence of the echo amplitude was
theoretically studied by Wurger, Fleischmann, and Enss
\cite{quadr,wurger} and by one of us \cite{diagram}. However, a
detailed comparison of the theoretical results with experiments was
not carried out in those works. This work is aimed at the detailed
comparison of the theoretical results \cite{diagram} with the
available experimental data. We perform a theoretical analysis of
the derived general formula in various limiting cases and show that
the theory agrees with the experimental data reasonably well, both
qualitatively and, in some cases, quantitatively.

\section{Interaction of two-level systems with nuclear quadrupole moments}

Let us briefly recall how the presence of atoms with nuclear
quadrupole moments affects the properties of two-level systems and,
in particular, the amplitude of the dipole echo in glasses
\cite{diagram}. Let one of the atoms that are displaced under the
tunneling of the TLS have a nuclear quadrupole moment. The energy of
its quadrupole interaction with the microscopic electric field and
Zeeman interaction with the external magnetic field leads to TLS
level splitting, so that the TLS becomes a system with two identical
level sets each containing $2J + 1$ levels. These sets are separated
by the energy E of the original TLS, which is much greater than the
energy splitting within the set.

The Hamiltonian of such a system may be written in the form
\cite{diagram}:
\begin{multline}
\label{H4} \widehat{\cal H}_{\rm tot} = \frac{1}{2} \left(
\begin{array}{cc}
E& 0\\
0& -E \\
\end{array}
\right) \otimes \widehat{1}_Q +
\widehat{1}_\sigma\otimes\widehat{\widetilde{W}}_Q +{}\\
{}+ (\mbox{\bf F}\cdot\mbox{\bf m})\,\frac{1}{E}\, \left(
\begin{array}{lc}
\Delta & \Delta_0\\
\Delta_0 &- \Delta \\
\end{array}
\right) \otimes\widehat{1}_Q
+{}\\
{}+ \frac{1}{E}\, \left(
\begin{array}{lc}
\Delta & \Delta_0\\
\Delta_0 &- \Delta \\
\end{array}\right)\otimes \widehat{\widetilde{V}}_Q
.
\end{multline}
Here, $\widehat{1}_\sigma$ and $\widehat{1}_Q$ are the $2 \times 2$
and $N \times N $ ($N = 2 J + 1$) unity matrices in the spaces of the
TLS and the projections of the nuclear spin, respectively. $\Delta$
is the difference between the double-well potential minima.
$\Delta_0$ is the tunneling amplitude of the initial (unsplit) TLS,
$E = \sqrt{\Delta^2+\Delta_0^2}$ is the total energy of the TLS
disregarding the fine structure effects. $\bf F$ is the ac electric
field, and $\bf m$ is the electric dipole moment of the TLS.
$\widehat{\widetilde{W}}_Q = {\bf \widehat{M}\cdot
H}+\widehat{Q}_{\alpha\beta}\varphi_{\alpha\beta}(0)$, where
$\widehat{\bf M}$ is the nuclear magnetic moment, $\bf H$ is the
external magnetic field, $\widehat{Q}_{\alpha\beta}$ is the operator
of the nuclear quadrupole moment \cite{Landau} (expressed in terms
of the nuclear spin operator), and $\varphi_{\alpha\beta}(0)$ is the
second-derivative tensor of the electric potential taken at the
double-well potential maximum.

The second term in Eq.~(\ref{H4}), which contains
$\widehat{\widetilde{W}}_Q$, causes the fine splitting of the TLS.
We assume that the basis nuclear spin wavefunctions are
eigenfunctions of the operator $\widehat{\widetilde{W}}_Q$ and its
matrix is diagonal in this basis. The last term in the Hamiltonian
(\ref{H4}) includes the operator $\widehat{\widetilde{V}}_Q =
\widehat{Q}_{\alpha\beta}\varphi_{\alpha\beta}^{\,'}(0)|x_{\rm
min}|/x_0$, where $\varphi_{\alpha\beta}^{\,'}(0)$ --- is the
derivative of the tensor $\varphi_{\alpha\beta}(x)$ with respect to the
generalized coordinate $x$ of the TLS at $x=0$ and $|x_{min}|/x_0$ is the
ratio of the atom displacement under the tunneling of the TLS to the
characteristic interatomic distance $x_0$.

The first term in Eq. (\ref{H4}) determines the two-level system,
the second one is responsible for the fine splitting of the TLS due
to the magnetic and quadrupole moments of the nucleus, and the third
term determines the matrix elements of the transitions between
various levels in the microwave field. Finally, the fourth term
couples the split levels of the system, which leads to allowed
transitions between different fine structure levels and, ultimately,
to the oscillations of the dipole echo in the magnetic filed. Since
$|x_{min}|/x_0 \ll 1$, the last term is small and may be taken into
account perturbatively.

Performing the calculations in the lowest (second) order of
perturbation theory, we come (similar to \cite{diagram}) to the
following expression for the echo amplitude:
\begin{multline}
P_{\rm echo} \propto
-i V_1 V_2^2 \left(\frac{\Delta_0}{E}\right)^4 \times {}\\
{}\times \left[ 1 - \frac{64}{N} \sum_{n,m>n}
\left(\frac{\Delta}{E}\right)^2
\left|(\widetilde{V}_Q)_{nm}\right|^2
\frac{\sin^4\left(\varepsilon_{nm}\tau/2\hbar\right)}{\varepsilon_{nm}^2}
\right]   . \label{gen}
\end{multline}
Here, $\tau$ is the time interval between two excitation pulses,
$\varepsilon_{nm} = (\widetilde{W}_Q)_{nn} - (\widetilde{W}_Q)_{mm}$
is the energy differences between the fine structure levels of the
TLS. Equation (\ref{gen}) differs from similar Eq.~(44) in
\cite{diagram} basically due to of the absolute-value
square $\left|(\widetilde{V}_Q)_{nm}\right|^2$, which implies that
our consideration is not limited by only the real-valued matrix
elements $(\widetilde{V}_Q)_{nm}$.

The magnetic field enters Eq. (\ref{gen}) first via the energy
difference $\varepsilon_{nm}$ and, second, implicitly via the matrix
elements $(\widetilde{V}_Q)_{nm}$, owing to the fact that they were
calculated in the basis in which $\widehat{\widetilde{W}}_Q$ is
diagonal.

According to our numerical calculations, the latter dependence is
usually insignificant for finding the echo amplitude for the nuclei
with integer spin $J$. However, when the spin is a half-integer and
the levels $n$ and $m$ are degenerate according to the Kramers
theorem, the matrix element $(\widetilde{V}_Q)_{nm}$ cannot remove
degeneracy and, consequently, must vanish in a zero magnetic field.

Formula (\ref{gen}) allows a numerical calculation of the echo
amplitude for given parameters of the system. However, a direct
application of this formula is complicated by the necessity of
solving the algebraic equation of degree $2J + 1$ with subsequent
averaging of the results (which cannot be done in the general form
even for $J = 1$). Thus, Eq.~(\ref{gen}) must be investigated in
various limiting cases and analyzed numerically.

\section{Analysis of limiting cases}

We begin the analysis of Eq. (\ref{gen}) with three independent
energy scales appearing in it. The first two are the Zeeman energy
$E_H$ and the energy $E_Q$ of the quadrupole interaction of the
nucleus with the microscopic electric field. The third scale $E_\tau
= \hbar/\tau$ is determined by the time interval $\tau$ between the
pulses.

When the magnetic field is so high that $E_H$ is greater than the
other two scales, we can replace
$\sin^4(\varepsilon_{nm}\tau/2\hbar)$ by its average value and
neglect the quadrupole interaction energy in the calculation of
$\widehat{W}_Q$ (and consequently $\varepsilon_{nm}$). In this case,
the echo amplitude turns out to be
\begin{equation}
\label{simple} P_{\rm echo} \propto 1- C/H^2 ,
\end{equation}
where the coefficient C is independent of the magnetic field.

Thus, the echo amplitude asymptotically (as $1/H^2$) approaches a
constant value with an increase in the magnetic field, which agrees
with the experimental data for high fields \cite{glyc,exp}. A less
trivial magnetic-field dependence of the echo amplitude takes place
when the Zeeman energy $E_H$ is comparable with (or less than) at
least one of the other energy scales $E_Q$ or $E_\tau$.

\subsection{Small quadrupole splitting}

Consider now the case where the quadrupole interaction energy is
$E_Q \ll E_H, E_\tau$. In this case, $E_H$ and $E_\tau$ may be
comparable. Then, we can neglect the quadrupole energy in the
calculation of $\widehat{\widetilde W}_Q$ (but we still have to
retain it in $\widehat{\widetilde V}_Q$). Correspondingly, the set
of the eigenfunctions of the projections of spin $\bf J$ on the
direction of the magnetic field can be used as the basis. In this
case, the fine structure levels are $E_n = \mu H (J-n-1)/J$, where
$\mu$ is the nuclear magnetic moment and $n = 1, 2,..., 2J + 1$.

In this case, since glass is isotropic and the trace of the tensor
$\varphi'_{\alpha\beta}(0)$ is zero, the average squares of the
absolute values of all matrix elements
$\left|(\widetilde{V}_Q)_{nm}\right|^2$ can be expressed in terms of
a single constant. Then, the dipole echo amplitude (as a function of
the magnetic field) is determined by the formula
\begin{multline}
\label{noQ} P_{\rm echo} \propto
-iV_1V_2^2 \left(\frac{\Delta_0}{E}\right)^4 \times{}\\
{}\times \left\{1 - C\, \frac{\Delta^2}{E^2}\left[ \frac{\sin^4(\mu
H\tau/2J\hbar)}{H^2} + \frac{\sin^4(\mu
H\tau/J\hbar)}{4H^2}\right]\right\} .
\end{multline}
Here, $C$ is the coefficient independent of the magnetic field (and
different from $C$ in Eq. (\ref{simple})).

The magnetic field enters into Eq. (\ref{noQ}) as the product $\mu
H\tau$. This implies that the variation of the dipole echo amplitude
with $H$ should scale as $1/\tau$ along the horizontal axis.

\begin{figure}[htbp]
    \centering
        \includegraphics[width=0.45\textwidth]{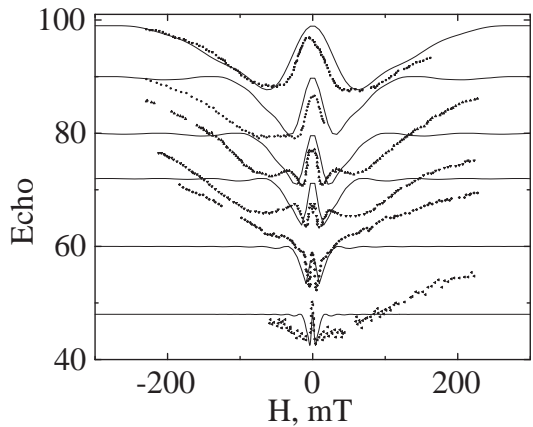}
        \caption{ Magnetic-field dependence of the dipole echo amplitude
        in the BK7 glass (in arbitrary units) for $\tau$ = (from top to bottom)
        0.75, 1.5, 2, 3, 6, and 12$\,\mu s$. The points are the experimental data
        from \cite{exp} and solid lines are the theoretical curves for the nuclear
        magnetic moment of $\rm ^{10}B$ calculated with the use of Eq.~(\ref{noQ}).
         }
    \label{fig:Graph1}
\end{figure}

Let us compare Eq.~(\ref{noQ}) with the measurement of the echo
amplitude in the BK7 glass as a function of the magnetic field~\cite{exp}
(Fig.~\ref{fig:Graph1}). Obviously, the experimental
dependence of the echo amplitude on H has two components, one of
which keeps its scale with a change in $\tau$ and the other one
changes the horizontal scale as $1/\tau$. The second component is
well described by Eq.~(\ref{noQ}) with $\mu$ equal to the nuclear
magnetic moment of the $\rm ^{10} B$ atom (which is present in the
BK7 glass and has a nuclear spin $J=3$). We may suggest that the
nonscalable component of the dependence is determined by the
contribution of other atoms (e.g., $\rm Na$ or $\rm ^{11} B$) with
the nuclear quadrupole moment, for which the condition $E_\tau \gg
E_Q$ is not fulfilled.

\subsection{Magnetic field dependence of echo amplitude in weak fields}

Consider now another limiting case where $E_Q \simeq E_\tau$ but the
Zeeman energy is low, $E_H \ll E_Q, E_\tau$. This will give us an
idea of the behavior of an echo amplitude in small fields.

\begin{figure}[htbp]
    \centering
        \includegraphics[width=0.47\textwidth]{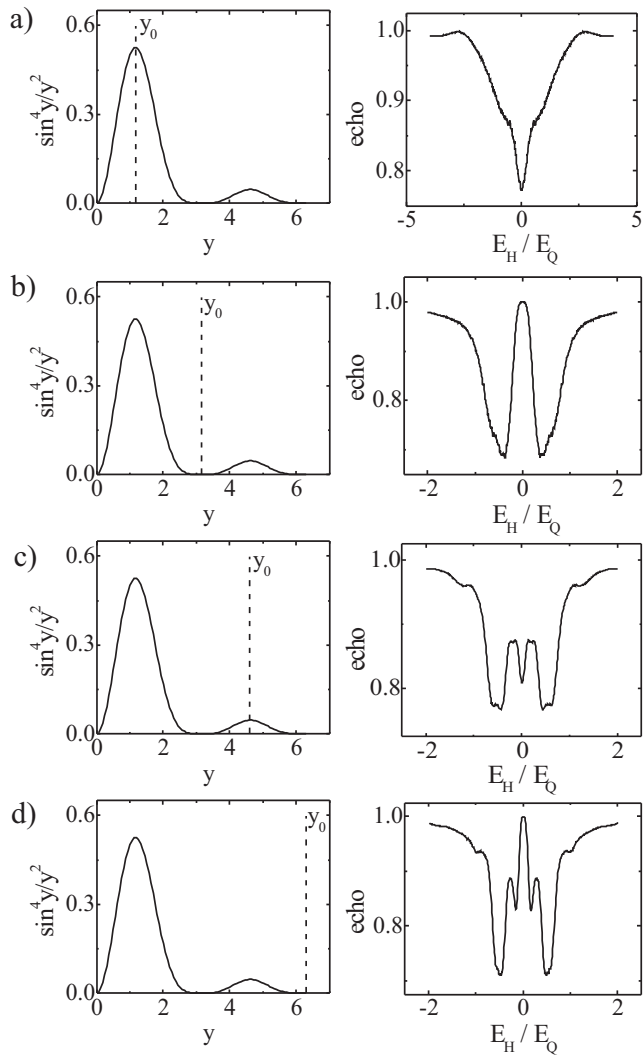}
        \caption{Positions of $y_0$ and the respective results of the simulation
        of the dipole echo amplitude for $J = 3/2$.
         }
\label{fig:cases}
\end{figure}

At $H = 0$, the energy differences $\varepsilon_{nm}$ between the
fine structure levels of the TLS are fully determined by the energy
of the quadrupole interaction of the nucleus with the internal field
and enter the final formula for the echo amplitude through the
function
\begin{equation}
\sin^4y/y^2 , \quad y=\varepsilon_{nm}\tau/2\hbar. \label{s4}
\end{equation}
The plot of this function (see Fig.~\ref{fig:cases}, left panes) is
a set of peaks with the $1/y^2$ envelope. The character of the
magnetic-field dependence of the dipole echo amplitude in weak
fields is determined by the fact that the values of $y_0$
corresponding to the zero-field splitting may fall to a minimum or
maximum of dependence (\ref{s4}).

Consider for example the simplest case of the nucleus with $J =
3/2$. This nucleus has four fine structure levels and, respectively,
three independent energy differences $\varepsilon_{nm}$. However,
two of them ($\varepsilon_{12}$ and $\varepsilon_{34}$) vanish in
zero magnetic field, according to the Kramers theorem. The
corresponding matrix elements $(\widetilde{V}_Q)_{nm}$ also vanish.
As a result, the contribution of these two pairs of levels appears
to be small compared to the contribution of the remaining
$\varepsilon_{23}$ pair.

Thus, in this case, the behavior of an echo in weak fields is
determined by the single energy difference $\varepsilon_{23}$.
Suppose that the time interval $\tau$ between the pulses is such
that $y_0 = \varepsilon_{23}(H = 0)\tau/2\hbar = \pi k$ with integer
$k$, which corresponds to a minimum of function (\ref{s4}). Then, in
the presence of the magnetic field, $y$ must shift from the minimum
position and, according to Eq.~(\ref{gen}), the echo amplitude must
decrease, irrespective of the properties of the system. Therefore,
the magnetic-field dependence of the dipole echo should have a
maximum at $H = 0$.

Figures \ref{fig:cases}b and \ref{fig:cases}d show the numerical
results for the echo amplitude calculated with the use of Eq.~(\ref{gen})
for the case of $J = 3/2$ and $y_0 = \pi$ and $2\pi$,
respectively (contributions of all levels were taken into account).
Clearly, in agreement with the prediction, the echo amplitude has a
maximum at zero magnetic field.

On the contrary, if $\tau$ is such that $y_0$ corresponds to a
maximum of function (\ref{s4}), then the echo amplitude must
increase in the presence of the magnetic field and the
magnetic-field dependence of an echo should have a minimum at $H =
0$. The corresponding numerical results are shown in Figs.
\ref{fig:cases}a and \ref{fig:cases}c.

For an arbitrary spin, this analysis is possible if the contribution
of one energy splitting $\varepsilon_{nm}$ prevails over the
contributions of the other pairs of levels. This occurs, e.g., if
the other $\varepsilon_{nm}$ values are large and can be disregarded
in Eq.~(\ref{gen}) due to the term $1/\varepsilon_{nm}^2$ .

Thus, it turns out that a zero-field maximum of the echo amplitude
should change to a minimum and vice versa with a change in the time
interval $\tau$ between the pulses.

\begin{figure}[htbp]
    \centering
        \includegraphics[width=0.47\textwidth]{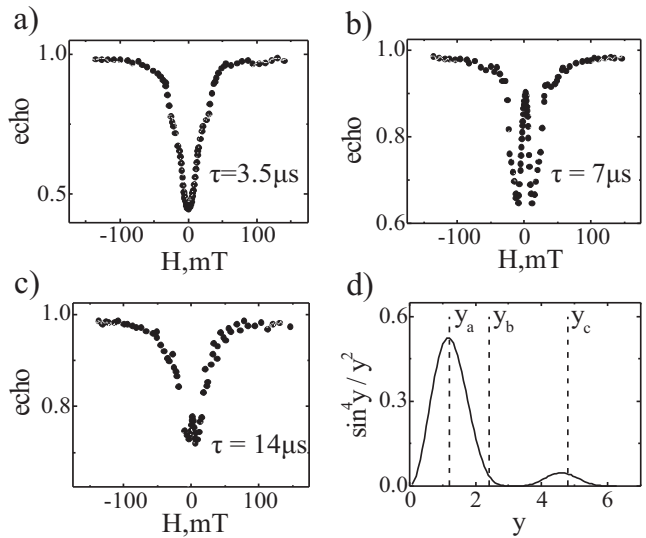}
        \caption{Experimental magnetic-field dependence of the
        dipole echo amplitude in glycerol~\cite{brandt,glycgraph} with $\tau$ = (a) $3.5$, (b) $7$,
        and (c) $14\,\mu s$ and (d) the respective zero-field positions $y_a$, $y_b$,
        and $y_c$ of $y_0$.}
        \label{fig:grex1}
\end{figure}

A similar behavior was experimentally observed in glycerol with
hydrogen replaced by deuterium~\cite{brandt,glycgraph}. Deuterium
has spin 1, which corresponds to three zerofield energy splittings
$\varepsilon_{12} \propto 2\eta$, $\varepsilon_{23}\propto 3 \eta$
and $\varepsilon_{13} \propto 3 + \eta$, where $0 < \eta < 1$ is the
parameter of the microscopic field asymmetry. If $\eta \ll 1$, the
first level splitting is much smaller than the other two and, in
agreement with the above analysis, may play the main role in the
magnetic-field dependence of the echo amplitude.

Figure 3 presents the experimental results~\cite{brandt,glycgraph}
and the positions of $y_0$ for different values of $\tau$. They were
calculated with the fitted value $\varepsilon_{12}/h \approx
110\,$\,kHz (which is comparable with an experimental value of
$150\,$\,kHz~\cite{glyc}). Clearly, the echo amplitude as a function
of the magnetic field has a minimum at $H = 0$ when $y_0$ appears
near the maximum of function (\ref{s4}) and vice versa; i.e., the
echo amplitude has a maximum at zero field when $y_0$ appears near
the minimum of the function (\ref{s4}).

Summarizing, the elaborated theory of the dipole echo in the
magnetic field with the inclusion of the nuclear quadrupole moments
provides the explanation of all characteristic features observed in
experiments and, in some cases, yield even a quantitative agreement
between the experiments and analytic calculations. However, the
comparison of the magnetic-field dependence of the dipole echo
amplitude in an arbitrary glass with the theory is complicated by
superimposed contributions from the atoms with different nuclear
magnetic and quadrupole moments.

We are grateful to A. Fleischmann and V.I. Kozub for fruitful
discussions. The work of A.V.S. in 2008 was supported by the
government of St. Petersburg (candidate's project no. 2.4/4-05/103).

\end{document}